\documentclass[aps, prd, twocolumn, showpacs, superscriptaddress, groupedaddress]{revtex4} 
\usepackage{graphicx}	
\usepackage{amssymb}
\usepackage{dcolumn}
\usepackage{color}
\usepackage{subfigure, rotating, bm, array}
\usepackage[pagebackref=false, colorlinks=true]{hyperref}
\hypersetup{
linkcolor=blue,     % color of internal links
citecolor=blue,     % color of links to bibliography
urlcolor=blue}      % color of the url
\begin{document}
\title{Shadow of a Naked Singularity without Photon Sphere}
\author{Ashok B. Joshi}
\email{gen.rel.joshi@gmail.com}
\affiliation{International Center for Cosmology, Charusat University, Anand 388421, Gujarat, India}
\author{Dipanjan Dey}
\email{dipanjandey.adm@charusat.edu.in}
\affiliation{International Center for Cosmology, Charusat University, Anand 388421, Gujarat, India}
\author{Pankaj S. Joshi}
\email{psjprovost@charusat.ac.in}
\affiliation{International Center for Cosmology, Charusat University, Anand 388421, Gujarat, India}
\author{Parth Bambhaniya}
\email{grcollapse@gmail.com}
\affiliation{International Center for Cosmology, Charusat University, Anand 388421, Gujarat, India}

\date{\today}

\begin{abstract}
It is generally believed that the shadows of either a black hole or naked singularity arise due to photon spheres developing in these spacetimes.  Here we propose a new spherically symmetric naked singularity spacetime solution of Einstein equations which has no photon sphere, and we show that the  singularity casts a shadow in the absence of the photon sphere. We discuss some novel features of this shadow and the lightlike geodesics in this spacetime. We compare the shadow of the naked singularity here with shadows cast by Schwarzschild black hole and the first type of Joshi-Malafarina-Narayan (JMN1) naked singularity, where for the last two spacetimes the shadow is formed due to the presence of a photon sphere. 
It is seen, in particular, that the size of shadow of the singularity is considerably smaller than that of a black hole. Our analysis shows that the shadow of this naked singularity is distinguishable from the shadow of a Schwarzschild black hole and the JMN1 naked singularity. These results are useful and important in the context of recent observations of shadow of the M87 galactic center.

%Here we cast shadow of the spherically symmetric, static naked singularity. The naked singularity is look likes Schwarzschild spacetime but it form shadow without photon sphere. Similar study for $M_0 = \frac{2}{3}$ in Joshi-Malafarina-Narayana(JMN)-1 spacetime. Schwarzschild spacetime form photon sphere  at $r = 3M$, while in this new spacetime photon sphere is at $r=0$ but in JMN-1 spacetime effective potential is become constant from $0$ to $R_b$. For the formation of shadow, b impact parameter play vital role. In Schwartzschild and JMN-1  spacetimes shadow form at $3\sqrt{3}M$ while for new spacetime at $M$. Recently photo of M87 object which can be reveal unknown fact about the central object, which can be naked singularity, black hole or compact object. So here we study, comparative study of naked singularity and black hole.
\bigskip

$\boldsymbol{key words}$ : Naked singularity spacetime, Black hole spacetime.
\end{abstract}
\maketitle
\section{Introduction}
It is well-known that when massive matter clouds undergo a catastrophic continual gravitational collapse, the total mass collapses into a spacetime singularity. At such a singularity, the  density, pressures and spacetime curvatures diverge \cite{spacesingularity2, spacesingularity1}. To understand the final state of such a collapse better, one could consider a collapsing spherically symmetric compact object within a dynamical spacetime configuration, where the internal dynamical metric seeded by the collapsing matter is matched  externally to a Schwarzschild solution. 

When the collapsing matter is homogeneous and dustlike, the final sate of the gravitational collapse is necessarily a Schwarzschild black hole. It then has a central spacetime singularity which is covered by a null hypersurface \cite{refn1}, known as the event horizon. Since this singularity is covered by a null surface, so no information from the same can travel to any other spacetime events or faraway observers. As this singularity is not causally connected to any other spacetime points, it is a spacelike singularity. 

On the other hand, there can be other types of singularity which are causally connected with other spacetime points. Such causally connected singularities are also known as naked singularities. There are two types of naked singularities:  nulllike and timelike singularities. One can show that both these singularities can be formed as the final state of gravitational collapse of an inhomogeneous matter cloud \cite{Joshi:1993zg,timesingularity,Kudoh:2000xs,strongnaked,psJoshi1,psJoshi2, dey1, dey2, dey3}. 
There are many well known spacetimes, e.g. Joshi-Malafarina-Narayan (JMN) spacetimes \cite{psJoshi1,psJoshi2}, Janis-Newman-Winicour (JNW) spacetime \cite{Janis:1968zz}, which have central timelike singularities. It can also be shown that the timelike and null  singularities formed during a gravitational collapse of massive matter cloud can be gravitationally strong \cite{Joshi:1993zg,timesingularity,Kudoh:2000xs,strongnaked,psJoshi1,psJoshi2, dey1, dey2, dey3}. Around a strong curvature naked singularity, the quantum gravitational effects should be very large and powerful. This may indicate and imply a possible quantum gravity resolution of strong spacetime singularity that may happen around the extreme curvature regions of a spacetime \cite{Goswami:2005fu, Szulc, Ashtekar}. Therefore, as timelike and null strong singularities are causally connected to the other external spacetime points, the quantum gravity effects can be locally or globally observable. 

In \cite{psJoshi1, psJoshi2}, it is shown that  JMN spacetimes with central timelike singularities are obtained as the asymptotic equilibrium state of a quasi-static gravitational collapse.
Also, it is important to investigate the observational signatures of timelike singularities  in the context of  recent observation of  shadow of the M87 galactic center by the Event Horizon Telescope (EHT) collaboration \cite{Akiyama:2019fyp}. As for our own galaxy, the trajectories of `S' stars around our galactic center as observed by the GRAVITY and SINFONI collaboration \cite{M87, Eisenhauer:2005cv, center1} are providing very useful data. Recently, many studies have been made \cite{dey1, geodesicjnw, bst1, bst2, bst3, Parth, Dipanjan, lensing1, lensing2, lensing3, rajbul1, rajbul2, accre1, accre2, collapse, grava1, grava2, grava3, Narayan, bambi}, where the behaviour of timelike and null geodesics around timelike singularities and other compact objects has been investigated. In \cite{rajbul1, accre1}, the shadows cast by JMN spacetimes are investigated and it is shown that the shadow cast by first type of JMN spacetime (JMN1) can be similar to the shadow cast by the Schwarzschild black hole. JNW naked singularity spacetime also can cast a shadow, and the properties of the shadow can be similar to the Schwarzschild black hole shadow \cite{accre1, accre2}. 

It is worth noting, however, that for JNW and JMN1 spacetimes, it is the existence of photon sphere that causes the shadow. On the other hand, in \cite{DRJ}, it is shown that a thin matter shell can also cast a shadow in the absence of a photon sphere in the spacetime. These results emphasize that the occurrence of a shadow is not 
a property of black holes only, but it also can be cast by other compact objects in the presence of a photon sphere or thin shell of matter \cite{rajbul1, accre1,DRJ}. 

In this paper, we present another new naked singularity solution of Einstein equations which resembles with a Schwarzschild spacetime at large distances. We show that this naked singularity  spacetime can cast and admit a shadow though it has no photon sphere, or any thin shell of matter. The interesting fact that emerges is that the shadow is created by the singularity itself. This is an intriguing feature, because just like a black hole or other compact objects, the singularity itself, when not covered by an event horizon, behaves like an `object' in its own right, even to the extent of casting a shadow. 
Our analysis here shows that the size of shadow of the singularity is considerably smaller than that of a black hole. It is seen that the shadow of this naked singularity is distinguishable from shadow of a Schwarzschild black hole and JMN1 naked singularity.
Since the shadow is cast by the singularity, one may possibly speculate that the effect of quantum gravitational resolution of the singularity could be likely identified from the observed shadow and the various detailed physical features it may exhibit. Also, this novel feature of the spacetime singularity could be important in the context of the recent observations of shadow of the galactic center
M87. 

The plan of the paper is as follows. In section (\ref{sec1}), we discuss  the JMN1 spacetime and the new asymptotically flat naked singularity spacetime is presented. In section (\ref{sec2}), we discuss the nature of light trajectories around the Schwarzschild black hole, JMN1 naked singularity, and the new naked singularity solution presented here. In that section, we also discuss the shadow of naked singularity, and compare the same with that of the black hole. Finally, in section (\ref{result}), we discuss our results. Throughout the paper we take $G=c=1$.

\section{Naked Singularity Spacetimes}\label{sec1}
In this section, we present a new spherically symmetric, asymptotically flat naked singularity spacetime. We first discuss the basic spacetime properties of this naked singularity spacetime, and then we briefly discuss the first type of Joshi-Malafaria-Narayan (JMN1) spacetime.
\bigskip

\subsection{The Naked singularity spacetime}
The line element of the proposed spherically symmetric, static naked singularity spacetime can be written as, 
\begin{equation}
ds^2 = -\frac{dt^2}{\left(1+\frac{M}{r}\right)^2}+\left(1+\frac{M}{r}\right)^2dr^2 +  +r^2d\Omega^2\,\, , 
\label{eq3}
\end{equation}
where, $d\Omega^2=d\theta^2+\sin^2\theta d\phi^2$ and $M$ is a positive constant, where we will prove that $M$ is the ADM mass of the above spacetime. The expression of the Kretschmann scalar and Ricci scalar for this spacetime are:
\begin{widetext}
\begin{eqnarray}
    R_{\alpha\beta\gamma\delta}R^{\alpha\beta\gamma\delta} &=& \frac{4M^2\left((M-2r)^2 r^4 + 4(M+r)^2 r^4 + (M+r)^4 (M+2r)^2\right)}{r^4 (M+r)^8}\,\, ,\\
    R &=& \frac{2M^3(M + 4r)}{r^2(M + r)^4}\,\, .
\end{eqnarray}
\end{widetext}
From the above expressions of the Kretschmann scalar and Ricci scalar it can be seen that the spacetime has a strong curvature singularity at the center $r=0$. No null surfaces such as an event horizon exist around the singularity in this spacetime. Therefore, the singularity is visible to the outside observer.  As we know, the ADM mass of a spacetime can be written as \cite{refn1, ADM},
%\begin{equation}
%    M_{ADM} = \frac{1}{16\pi} \lim_{r \to \infty} \sum_{\mu, \nu = 1}^{3} \int \left(\frac{\partial h_{\mu \nu}}{\partial x^\mu} - \frac{\partial h_{\mu \mu}}{\partial x^\nu}\right) N^\nu dA \,\, ,
%\end{equation}
\begin{equation}
    M_{ADM} = - \frac{1}{8\pi} \lim_{S \to \infty} \oint_S \left(\mathbb{R}-\mathbb{R}_0\right) \sqrt{\sigma}  d^2\theta\,\, ,
\label{ADM1}
\end{equation}
where $S$ is the bounded two-surface. $\mathbb{R}$ is the extrinsic curvature of the two-surface $S$ embedded in spacelike hypersurface $\Sigma$, $\mathbb{R}_0$ is the extrinsic curvature of S embedded in flat space and the infinitesimal area of the two-surface is written as $\sqrt{\sigma}d^2\theta = r^2\sin\theta d\theta d\phi$, where $\sigma$ is the determenant of the induced metric on the two-surface. The limit $S\to\infty$ implies that the radius of the two-surface $S$ tends to infinity. For the above spacetime (eq.~(\ref{eq3})), the unit normal on the $S$ can be written as,
\begin{equation}
    n_{\alpha} = \sqrt{g_{rr}}~\partial_{\alpha}r=\left(1+\frac{M}{r}\right) \partial_{\alpha}r\,\, .
\end{equation}
The extrinsic curvature of $S$ embedded in $\Sigma$ can be calculated as,
\begin{equation}
    \mathbb{R} = n^{\alpha}_{;\alpha} =\frac{2}{r\left(1+\frac{M}{r}\right)}\,\, ,
\end{equation}
and the extrinsic curvature of $S$ embedded in flat space is $\mathbb{R}_0 = \frac{2}{r}$. Therefore, using eq.~(\ref{ADM1}), the ADM mass of the new naked singularity spacetime (eq.~(\ref{eq3})) can be written as,
\begin{equation}
    M_{ADM}= -\frac{1}{2} \lim_{r \to \infty}r^2\left[\frac{2}{r\left(1+\frac{M}{r}\right)}-\frac2r\right]=M\,\, .
\end{equation}
Hence, the ADM mass of the proposed naked singularity spacetime is $M$.
The above metric (eq.~(\ref{eq3})) resembles with Schwarzschild metric at a large distance. Therefore, at a sufficient distance from the naked singularity, the behaviour of timelike and null geodesics are similar to what is seen in the Schwarzscild geometry. However, near the singularity, the causal structure of this spacetime becomes different from the causal structure of Schwarzscild spacetime. Distinguishable behaviour of null geodesics and the shadow cast by the naked singularity in tnis spacetime are discussed in the next section.

Using Einstein field equations,  we can write the energy density and pressures of this naked singularity spacetime as:
\begin{equation}
    -T^t _t = \rho = \frac{M^2 \left(M + 3r\right)}{\kappa r^2 \left(M + r\right)^3}\,\, , \label{eq10}
\end{equation}
\begin{equation}
    T^r _r = p_r = -\frac{M^2 \left(M + 3r\right)}{\kappa r^2 \left(M + r\right)^3}\,\, , \label{eq11}
\end{equation}
\begin{equation}
    T^\theta _\theta = T^\phi _\phi = p_\theta = p_\phi = \frac{3M^2}{\kappa r^4} \left(1 + \frac{M}{r}\right)^{-4}\,\, . \label{eq12}
\end{equation}
To satisfy strong, weak and null energy conditions, we need
\begin{eqnarray}
   \rho &=& \frac{M^2 \left(M + 3r\right)}{\kappa r^2 \left(M + r\right)^3} \geq 0\,\, ,\\
   \rho + p_r &\geq & 0,\\
   \rho + p_\theta &=& \frac{M^2 \left(M^2 + 4Mr + 6r^2\right)}{\kappa r^2 \left(M + r\right)^4} \geq 0,\\
   \rho + p_r + 2p_\theta &=& \frac{6M^2}{\kappa r^4} \left(1 + \frac{M}{r}\right)^{-4} \geq 0\,\, ,
\end{eqnarray}
and it can be easily verified that all these conditions are fulfilled for this spacetime. As it can be seen, this spacetime is seeded by an anisotropic fluid, where 
the anisotropy in the pressures can be written as:
\begin{equation}
    p_r - p_\theta =- \frac{M^2 \left(M^2 + 4Mr
+6r^2\right)}{\kappa r^2 \left(M + r\right)^4}
\end{equation}
The equation of state ($\alpha$) for an anisotropic fluid can be written as \cite{deyd1}:
\begin{eqnarray}
 \alpha = \frac{2p_\theta + p_r}{3\rho} \,\, .
\end{eqnarray}
Therefore, using Equ. (\ref{eq10}), (\ref{eq11}) and (\ref{eq12}), we can write the equation of state for this spacetime as, 
\begin{equation}
    \alpha = \frac{2}{\left(3+\frac{M}{r}\right) \left(1+\frac{M}{r}\right)} -\frac{1}{3}\,\, ,
\end{equation}
where we can clearly see that as $r$ tends to zero, equation of state becomes $-1/3$. On the other hand, as $r$ tends to infinity, equation of state becomes $+1/3$. It is important to know the source of the anisotropic fluid which seeds such a particular spacetime. As we know,
to be a minimally coupled static scalar field solution \cite{deyd1} of Einstein equations, a spacetime should satisfy :
\begin{equation}
    R^{\mu\nu}-\frac12 g^{\mu\nu}R = \kappa \left( g^{\mu\nu} L + 2 \frac{\partial L}{\partial g_{\mu\nu}}\right)\,\, ,
\end{equation}
where $L = \left(-\frac{1}{2} g^{\alpha\beta} \partial_\alpha \Phi \partial_\beta \Phi - V(\Phi)\right)$ and $\Phi$ is scalar field which satisfy $\Box\Phi(r) = V^{\prime}(\Phi(r))$, where $V(\Phi)$ is the scalar field potential. Using the above equations, one can check that for minimally coupled scalar field solution, we need $\rho + p_\theta = 0$. From eqs.~(\ref{eq10},\ref{eq11},\ref{eq12}), it is easy to verify that this naked singularity spacetime is not a minimally coupled scalar field solution of Einstein equations.

\subsection{First type of Joshi-Malafarina-Narayan (JMN1) Spacetime} The line eliment of the JMN1 spacetime can be written as,
\begin{eqnarray}
 ds^2=-(1-M_0) \left(\frac{r}{R_b}\right)^\frac{M_0}{1-M_0}dt^2 + \frac{dr^2}{1-M_0} + r^2d\Omega^2\,\, , 
\label{JMN-1metric}
\end{eqnarray}
where the dimensionless quantity $M_0$ should be $0<M_0<1$ and at $R_b$, JMN1 spacetime matches with the external Schwarzschild spacetime. In \cite{psJoshi1}, it is shown that in asymptotic time, JMN1 spacetime can be formed as an endstate of the gravitational collapse of anisotropic matter fluid. This spacetime has a timelike strong singularity at the center \cite{psJoshi1}. Using the parameters of JMN1 spacetime, the external Schwarzschild spacetime can be written as,
\begin{eqnarray}
 ds^2=-\left(1-\frac{M_0 R_b}{r}\right)dt^2 + \frac{dr^2}{\left(1-\frac{M_0 R_b}{r}\right)} + r^2d\Omega^2\,\, .
\label{exSCH}\nonumber
\end{eqnarray} 
One can prove that this spacetime smoothly matches with external Schwarzschild metric.
JMN1 spacetime is an anisotropic fluid solution of Einstein equations, where the energy density and pressures can be written as,
\begin{eqnarray}
\rho = \frac{M_0}{r^2},\,\, P_r = 0,\,\, P_\theta = \frac{M_0^2}{4(1-M_0)r^2}\,\, .
\end{eqnarray} 
Therefore, the equation of state for this spacetime can be written as, $\alpha = \frac{M_0}{6(1-M_0)}$. There are many literature where particle trajectories in JMN1 spacetime and the shadow cast by the same is discussed, and it is shown that those results are very important to understand the distinguishable observational signatures of black holes and naked singularities.  

In the next section, we discuss about the properties of lightlike geodesics and the shadow of naked singularity mentioned in eq.~(\ref{eq3}).
\begin{figure*}
\centering
\subfigure[Figure of effective potetial in JMN1 spacetime.]
{\includegraphics[width=76mm]{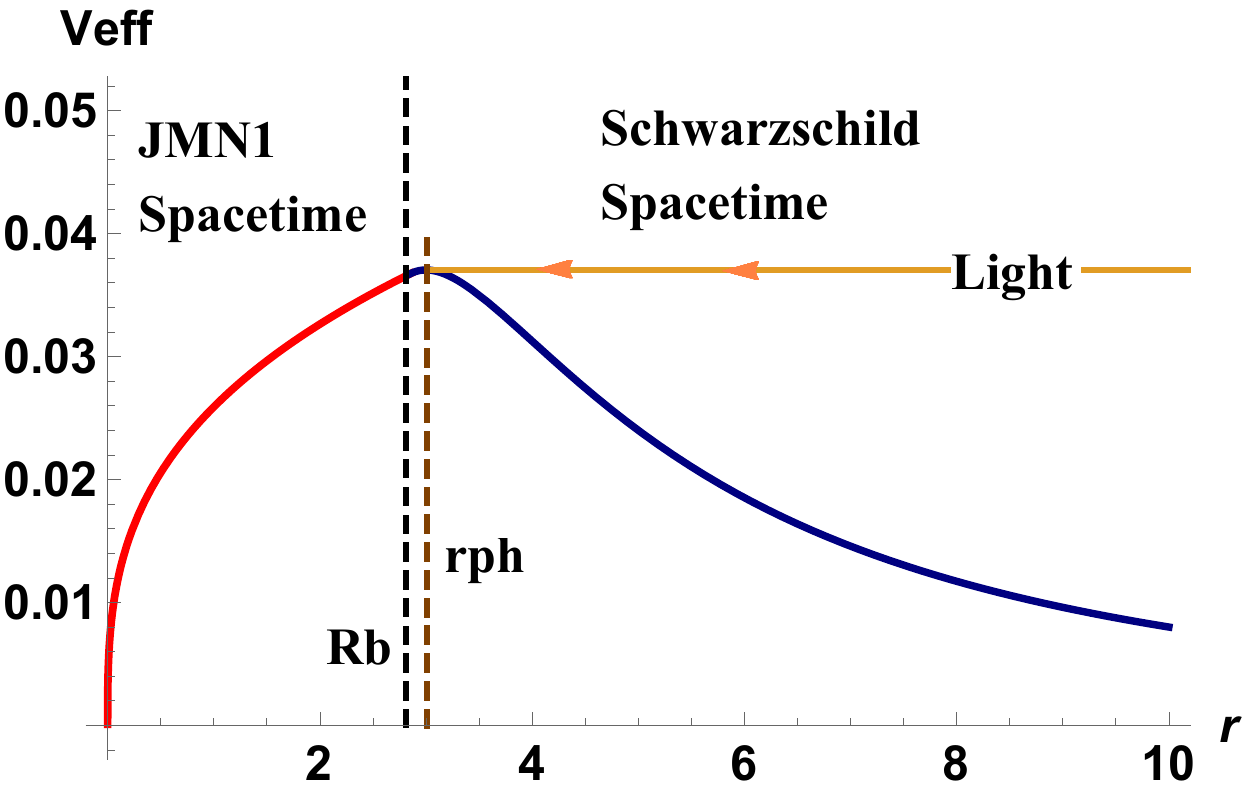}\label{fig1re1}}
\hspace{0.2cm}
\subfigure[Lensing effect in JMN1 spacetime.]
{\includegraphics[width=67mm]{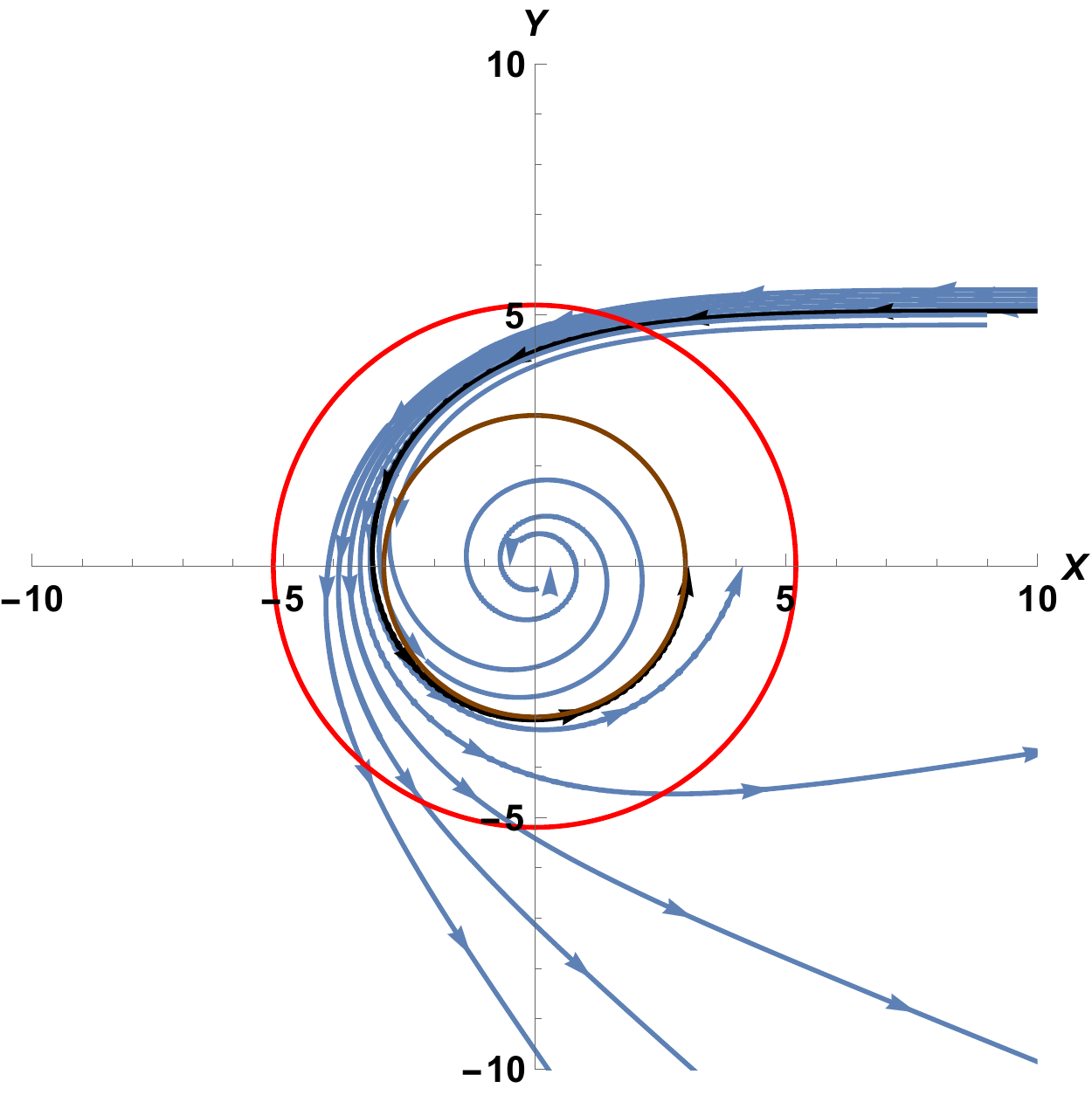}\label{fig1re2}}\\
\hspace{0.2cm}
\subfigure[Figure of effective potential in Schwartzschild spacetime.]
{\includegraphics[width=67mm]{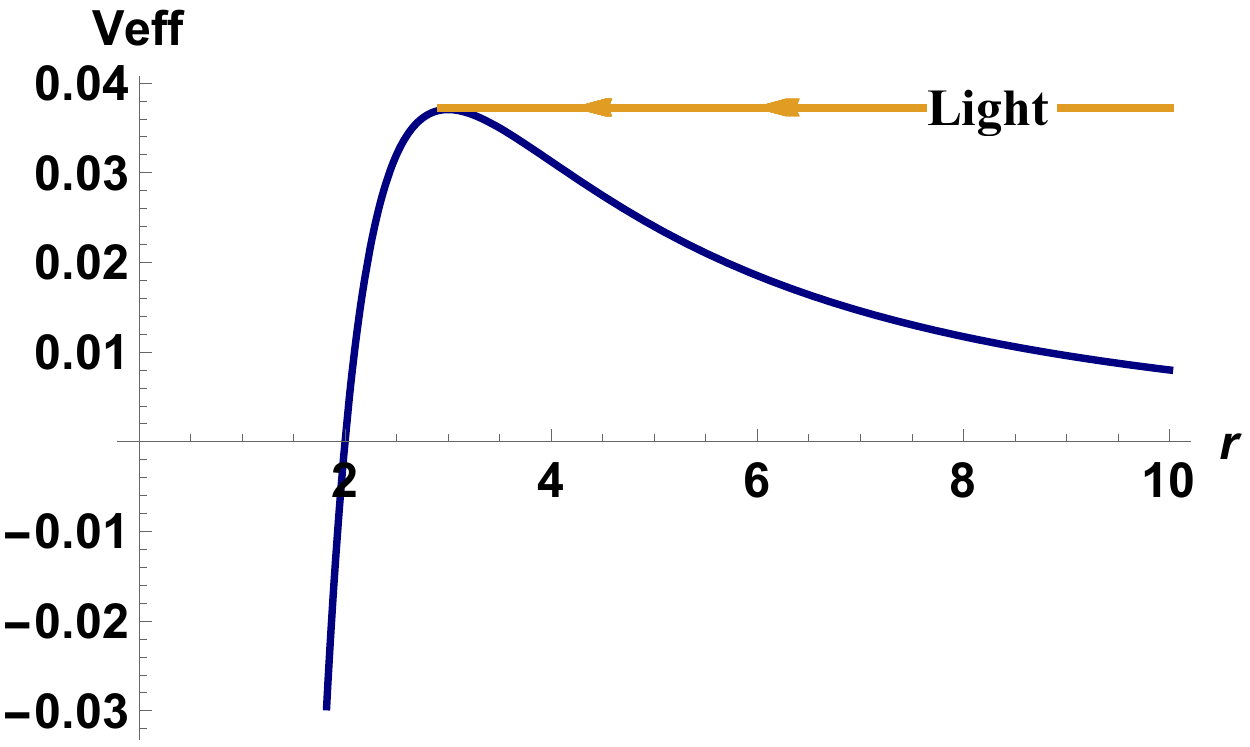}\label{fig1re3}}
\hspace{0.2cm}
\subfigure[Lensing effect in Schwartzschild spacetime.]
{\includegraphics[width=67mm]{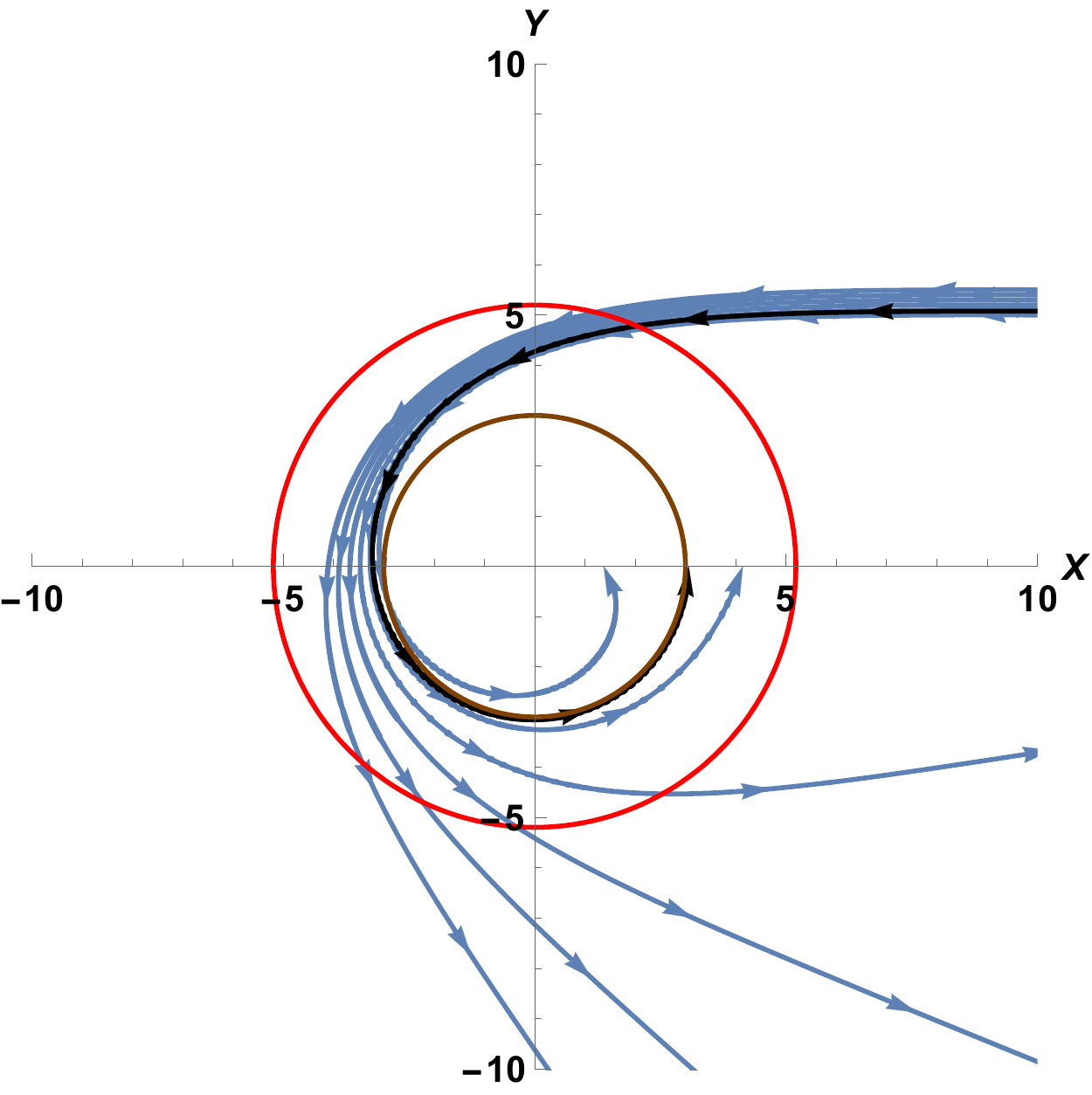}\label{fig1re4}}\\
\subfigure[Figure of effective potential in new spacetime.]
{\includegraphics[width=67mm]{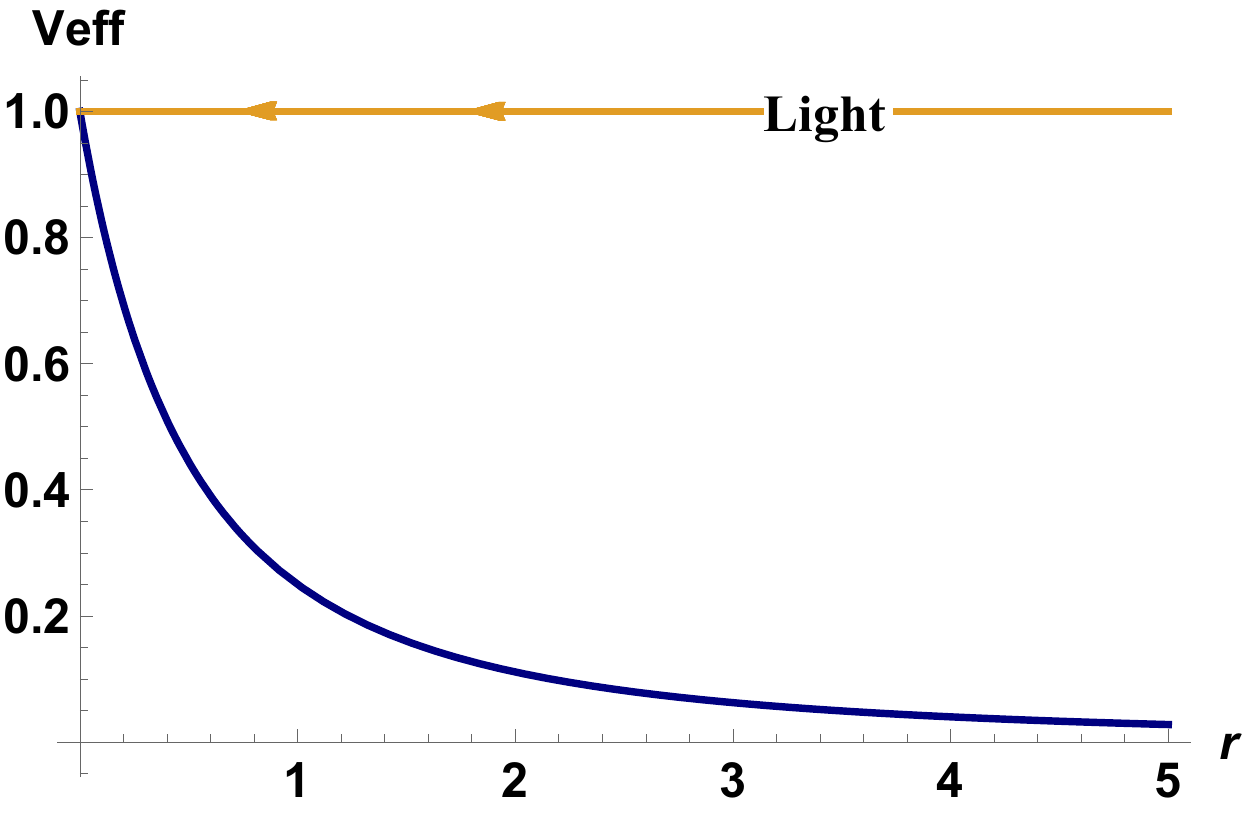}\label{fig1re5}}
\hspace{0.2cm}
\subfigure[Lensing effect in new spacetime.]
{\includegraphics[width=67mm]{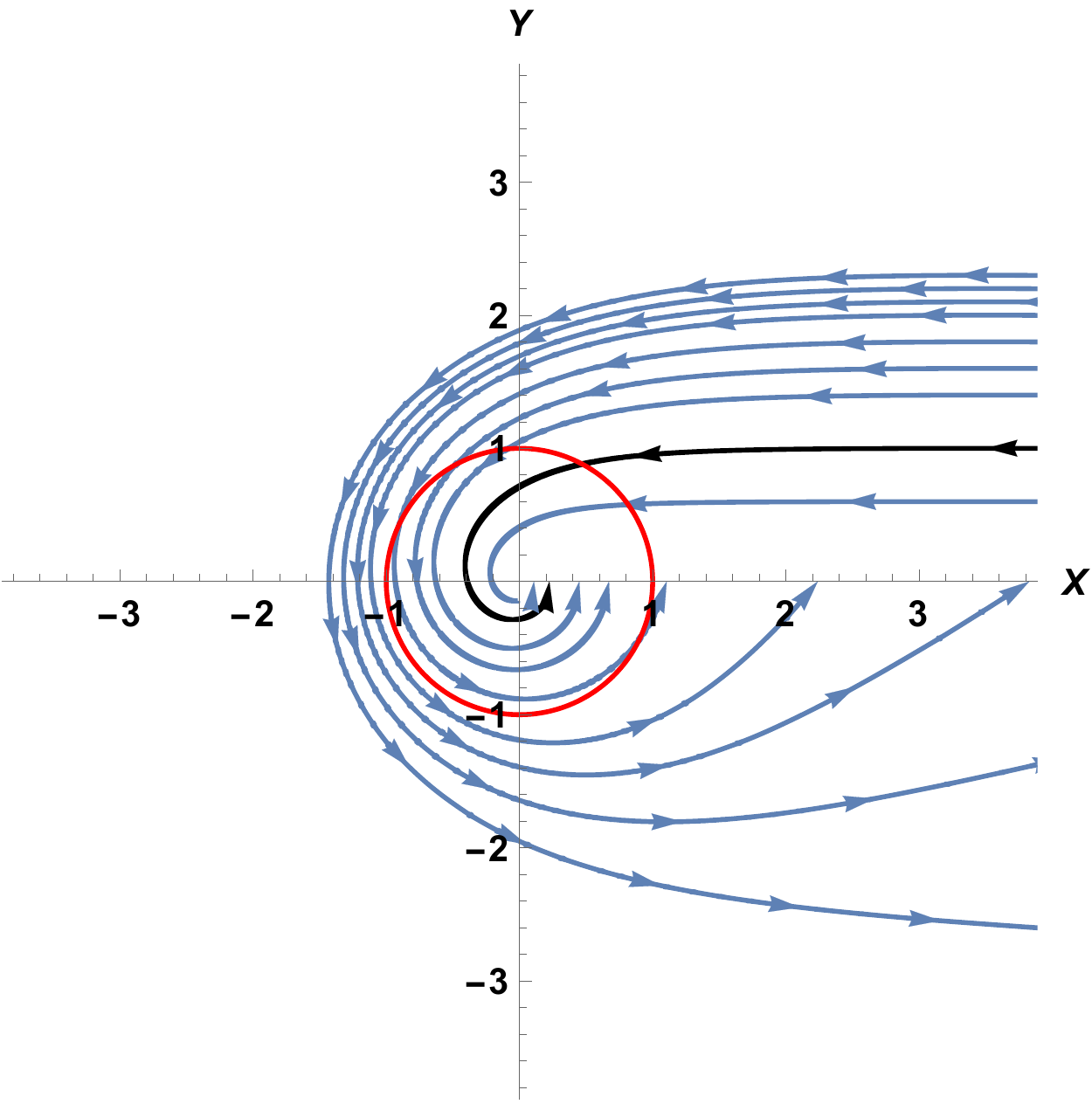}\label{fig1re6}}\\
 \caption{In figs.~(\ref{fig1re1}), (\ref{fig1re3}), (\ref{fig1re5}) , we show the nature of  effective potentials of the null geodesics in JMN1, Schwarzschild and the naked singularity spacetime
 given here respectively. In figs.~(\ref{fig1re2}), (\ref{fig1re4}), (\ref{fig1re6}), the light trajectories in these spacetimes are shown. In figs.~(\ref{fig1re2}), (\ref{fig1re4}), (\ref{fig1re6}), the blue lines are the null geodesics. Effective potential of lightlike geodesics in JMN1 spacetime configuration is shown in fig.~(\ref{fig1re1}). As we can see, the effective potential has a maximum value in JMN1 spacetime configuration and in Schwarzschild spacetime. Therefore, for these two spacetimes, photon spheres exist. The brown circles in figs.~(\ref{fig1re2}), (\ref{fig1re4}), show the size of the photon sphere. The red circles shows the size of the shadow.  Detailed analysis of these figures is given in the text.}\label{fig1}
\end{figure*}

\section{Shadows of Naked singularity and Black hole spacetimes}\label{sec2}
A spherically symmetric, static spacetime can be written as, 
\begin{equation}
    ds^2 = - g_{tt}dt^2 + g_{rr}dr^2 + r^2(d\theta^2 + \sin^2\theta d\phi^2)\,\,,
    \label{static}
\end{equation}
where $g_{tt}$,~$g_{rr}$ are the functions of $r$ only and the azimuthal part of the spacetime shows the spherical symmetry. For the lightlike geodesics in the above spacetime, in the $\theta=\frac{\pi}{2}$ plane, we can write 
\begin{equation}
    \frac{1}{b^2}=\frac{g_{tt} g_{rr}}{h^2} \left(\frac{dr}{d\lambda}\right)^2 + V_{eff}\,\, ,
\end{equation}
where the effective potential $V_{eff} = g_{tt}/r^2$ and the impact parameter $b = \frac{h}{\gamma}$, where $\gamma$ and $h$ are conserved energy and conserved angular momentum per unit rest mass respectively. In the above equation, we use $k_{\mu}k^{\mu}=0$, where $k^{\mu}$ is the null four- velocity. From the nature of effective potential of photons, one can investigate the stable and unstable orbits of photons. 
As we know, when this effective potential has a maximum value at some radius, we get unstable circular lightlike geodesics. The sphere corresponding to the radius of these unstable circular lightlike geodesics is known as the photon sphere. At the photon sphere radius ($r_{ph}$), we can write $V_{eff}(r_{ph})=\frac{\gamma^2}{h^2}$, $V^{\prime}_{eff}(r_{ph})=0$ and $V^{\prime\prime}_{eff}(r_{ph})<0$. 

When these conditions are satisfied in a spacetime, we can say that there would exist a photon sphere at $r_{ph}$. For Schwarzschild spacetime, photon sphere exists at $r_{ph}=3M_{T}$, where $M_{T}$ is the Schwarzschild mass. The turning point ($r_{tp}$) of lightlike geodesics can be found when $V_{eff}(r_{tp})=\frac{\gamma^2}{h^2}=\frac{1}{b_{tp}^2}$. Therefore, from the expression of $V_{eff}$, at the turning point, we can write the impact parameter $b_{tp}$ as, 
\begin{equation}
    b_{tp} = \frac{r_{tp}}{\sqrt{g_{tt}(r_{tp})}}\,\, .
    \label{impact1}
\end{equation}
In a spacetime, when only one extremum value of effective potential of null geodesics exists and if the extremum value is the maximum value of the potential, then the minimum impact parameter of turning point of light geodesics is $b_{tp}= b_{ph}$, where $b_{ph}$ is the impact parameter corresponding to the photon sphere. Lightlike geodesics, coming from faraway source,  cannot reach to the faraway observer with the impact parameter $b< b_{ph}$. They will be trapped inside the photon sphere. Therefore, the light geodesics coming from behind the lens with impact parameter $b< b_{ph}$, cannot reach to the outside observer and this creates a shadow of radius $b_{ph}$ in the observer's sky. Hence, one can conclude that in this case, it is the photon sphere's shadow which can be seen in observer's sky.  On the other hand, when the photon sphere does not exist in a spacetime and the effective potential diverges at origin, then such a shadow would not form in that spacetime. 

In fig.~(\ref{fig1}), we show the trajectory of photons which are coming out from the far away source and deflected due to the curvature in JMN1, Schwarzschild and the new naked singularity spacetimes. In JMN1 spacetime, photon sphere exists when $M_0> \frac{2}{3}$ and for $M_0<\frac23$, photon sphere does not exist. For $M_0> \frac23$, JMN1 spacetime matches with external Schwarzschild metric at $R_b< 3M_T$, where $M_T$ is the Schwarzschild mass. Therefore, effectively, the spacetime configuration has a photon sphere which is in the external Schwarzschild spacetime. However, inside the photon sphere, at the center, there exists a spacetime singularity which is not covered by any event horizon. Therefore, around the photon sphere lightlike geodesics behave in the similar way what can be seen around the photon sphere of a black hole. The effective potential and the behaviour of lightlike geodesics in JMN1 spacetime are shown in fig.~(\ref{fig1re1}) and fig.~(\ref{fig1re2}) respectively, where we take Schwarzschild mass $M_T=1$ and $M_0=0.7$. Therefore, the JMN1 spacetime matches with external Schwarzschild spacetime at $R_b=2.857$ and the radius of the photon sphere corresponding to the external Schwarzschild spacetime is $r_{ph} = 3$. Hence, for $M_0>\frac23$, shadow is effectively cast by the photon sphere of the external Schwarzschild spacetime. In fig.~(\ref{fig1re3}) and fig.~(\ref{fig1re4}), we show the effective potential and nature of light geodesics respectively in the Schwarzschild spacetime  with Schwarzschild mass $M_T = 1$. The brown circles in fig.~(\ref{fig1re2}) and fig.~(\ref{fig1re4}) show the size of photon sphere ($r_{ph}=3$) in JMN1 spacetime configuration and in Schwarzschild spacetime respectively and the red circles in these two figures, show the size of the impact parameter ($b_{ph}$) corresponding to the radius of the photon sphere. From the eq.~(\ref{impact1}), one can calculate that the shadow radius or the impact parameter corresponding to the photon sphere radius is $b_{ph}= 3\sqrt{3} M_T= 3\sqrt3$.

For the new naked singularity spacetime, the minimum turning point radius is $r_{tp}=0$. In fig.~(\ref{fig1re5}), it is shown that the effective potential of lightlike geodesics has a finite value. Therefore, photons with finite energy can reach very close to the central naked singularity. Photons having energy $E_{ph}\geq V_{eff}(0)$, can reach to the central singularity. Now, the value of effective potential and the impact parameter corresponding to the minimum value of the radius of the turning point (i.e. $r_{tp}=0$) are,
 \begin{equation}
     V_{eff}(r_{tp})|_{r_{tp}=0}=\frac{1}{(r_{tp}+M)^2}|_{r_{tp}=0}=\frac{1}{M^2}\,\, ,
 \end{equation}
 \begin{equation}
     b_{tp}|_{r_{tp}=0}=(r_{tp}+M)|_{r_{tp}=0}=M\,\, .
 \end{equation}
Therefore, we can get finite value of the impact parameter corresponding to the zero value of minimum turning point radius and that finite value of the impact parameter $b_{tp}(r_{tp}=0)=M$, where $M$ is the ADM mass of the naked singularity spacetime given here. Hence, though the value of the minimum turning point radius is zero, due to the finite value of corresponding impact parameter, in this case also, the lightlike geodesics coming from a distant source with impact parameter $b<M$, must be trapped closer to the singularity. Therefore, in this naked singularity spacetime, the curvature of the spacetime itself around the singularity, can cast a shadow in the observer sky, even when there exists no photon sphere in this case. As the minimum turning point of the photons is at the singularity, one can conclude that it is the shadow of the singularity itself. In fig.~(\ref{fig1re6}), we show the nature of lightlike geodesics in this  naked singularity spacetime. In that figure, the lightlike geodesic highlighted in black corresponds to the critical impact parameter $b_{tp}=M$. From figs.~(\ref{fig1re2}), (\ref{fig1re4}) and (\ref{fig1re6}), the distinguishable nature of lightlike geodesics in the new naked singularity spacetime, and in JMN1 and Schwarzschild spacetimes, can be clearly seen. We note that to visualize the shadow of a central massive object which can be seen in a physical situation, one needs to calculate the intensity of light coming out from the accreting matter around the massive central object. 
 \begin{figure*}
\centering
\subfigure[Intensity distribution in Schwartzschild spacetime.]
{\includegraphics[width=82mm]{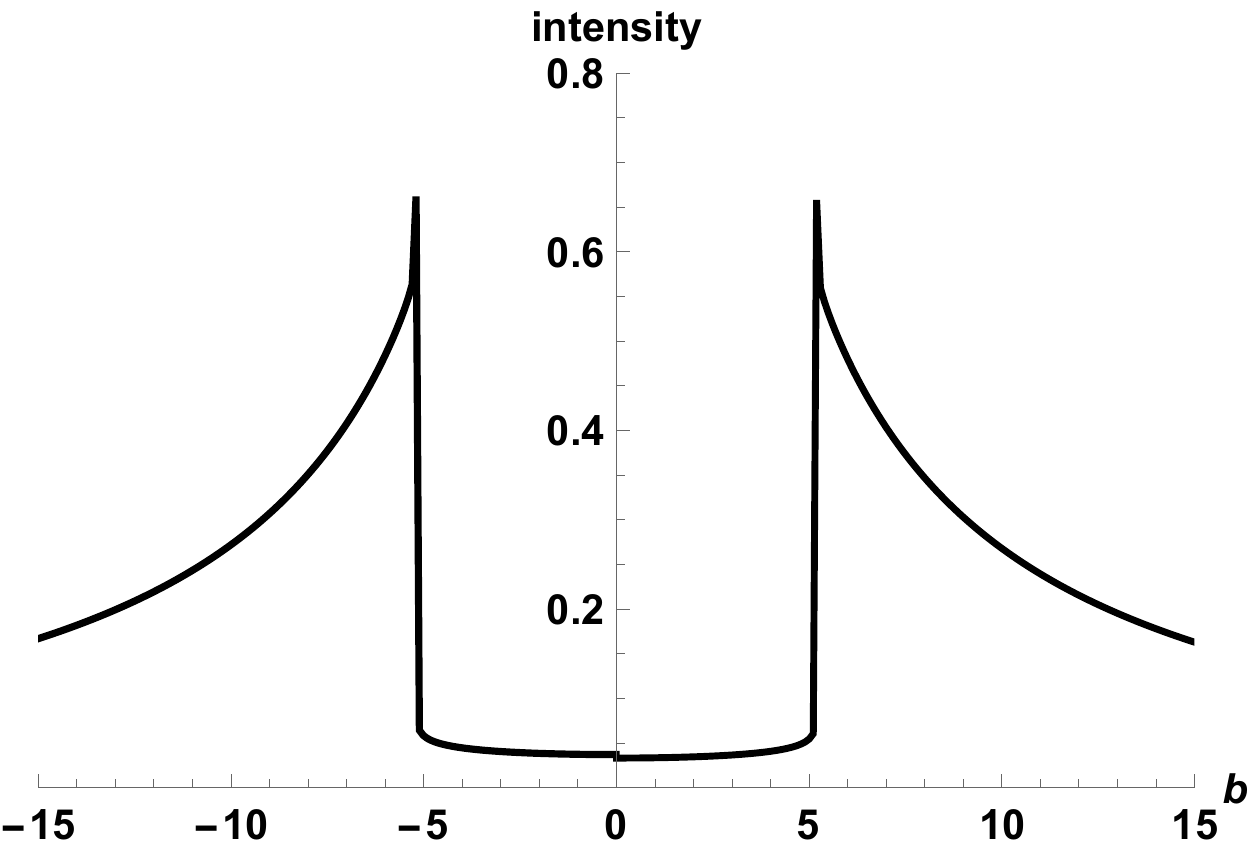}\label{re11}}
\hspace{0.2cm}
\subfigure[Shadow in Schwartzschild spacetime.]
{\includegraphics[width=68mm]{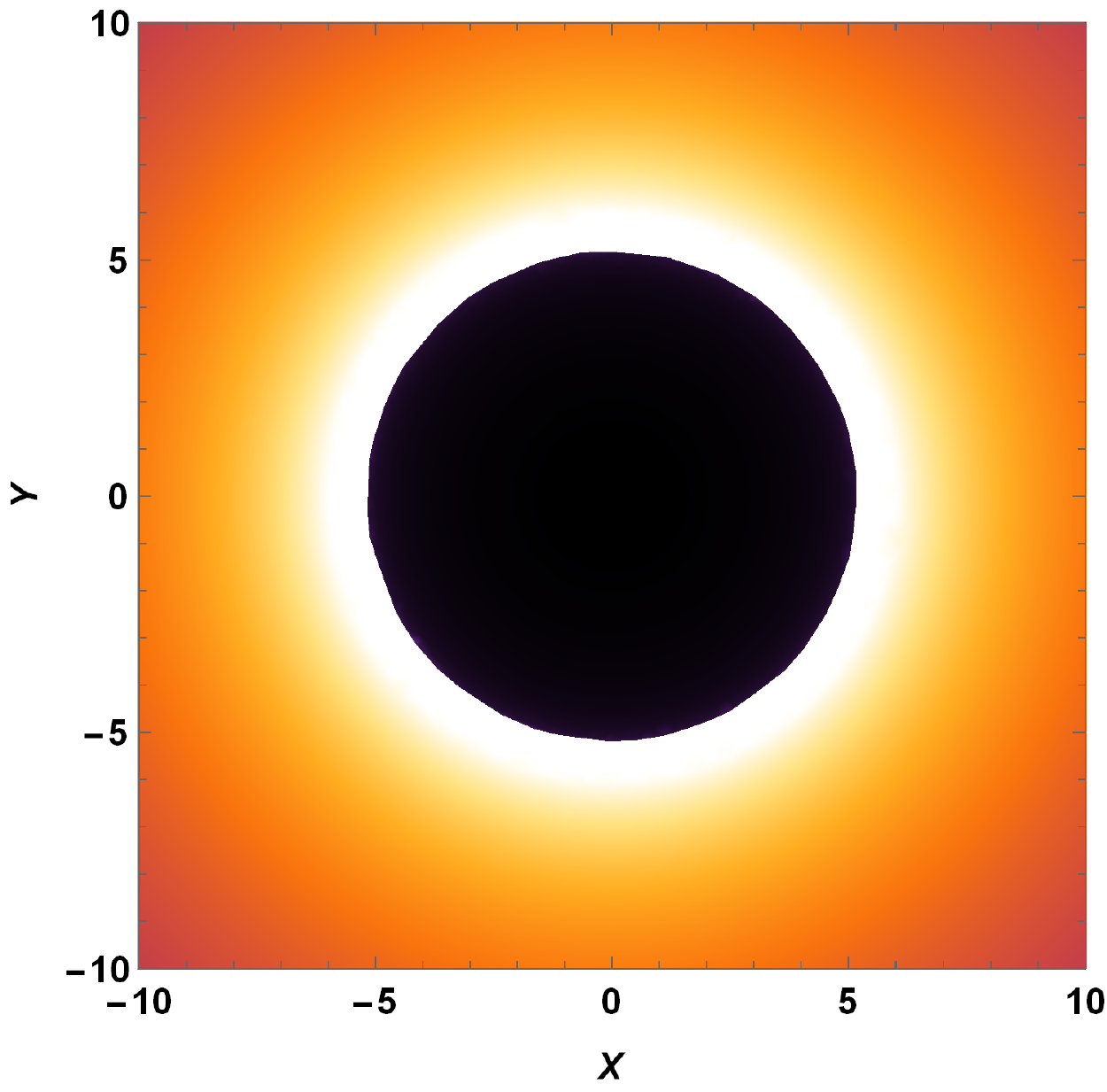}\label{re21}}\\
\subfigure[Intensity distribution in JMN-1 spacetime.]
{\includegraphics[width=82mm]{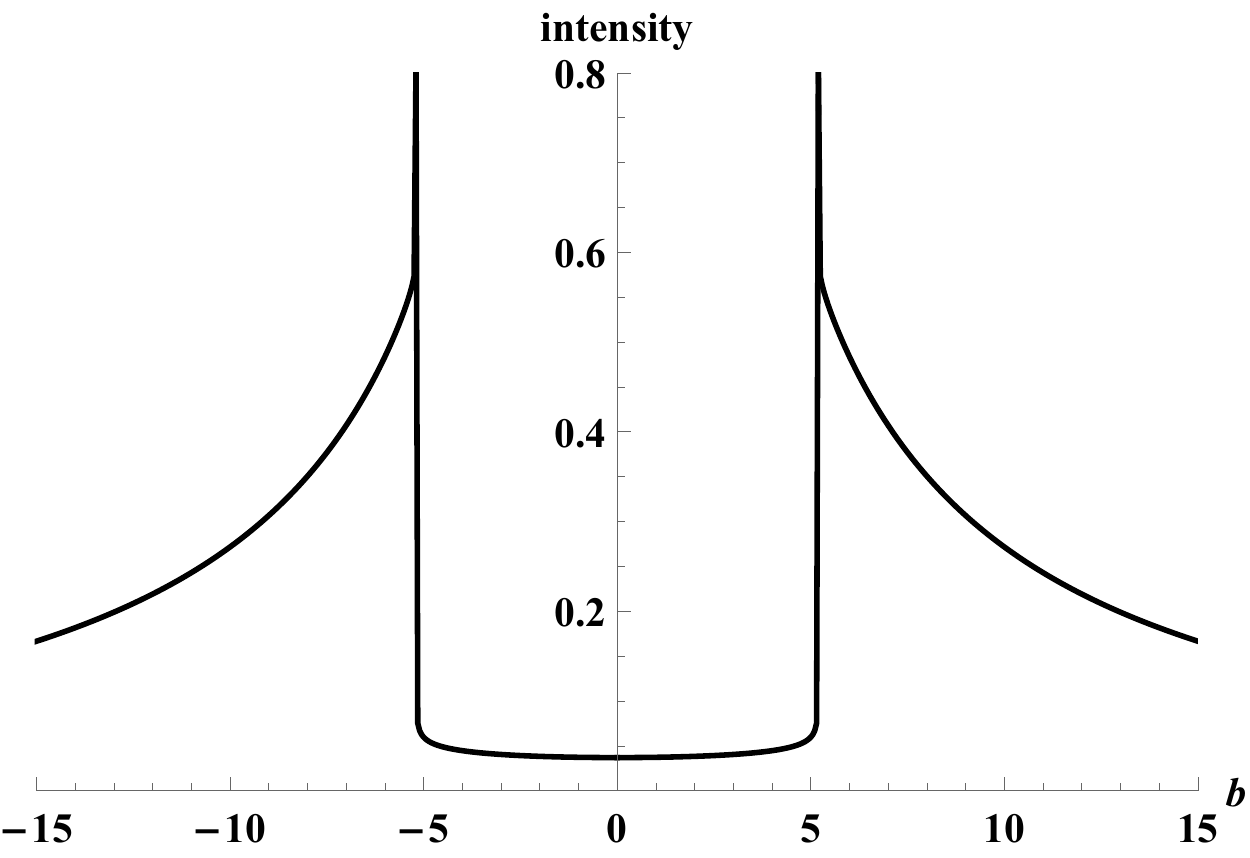}\label{re31}}
\hspace{0.2cm}
\subfigure[Shadow in JMN-1 spacetime.]
{\includegraphics[width=68mm]{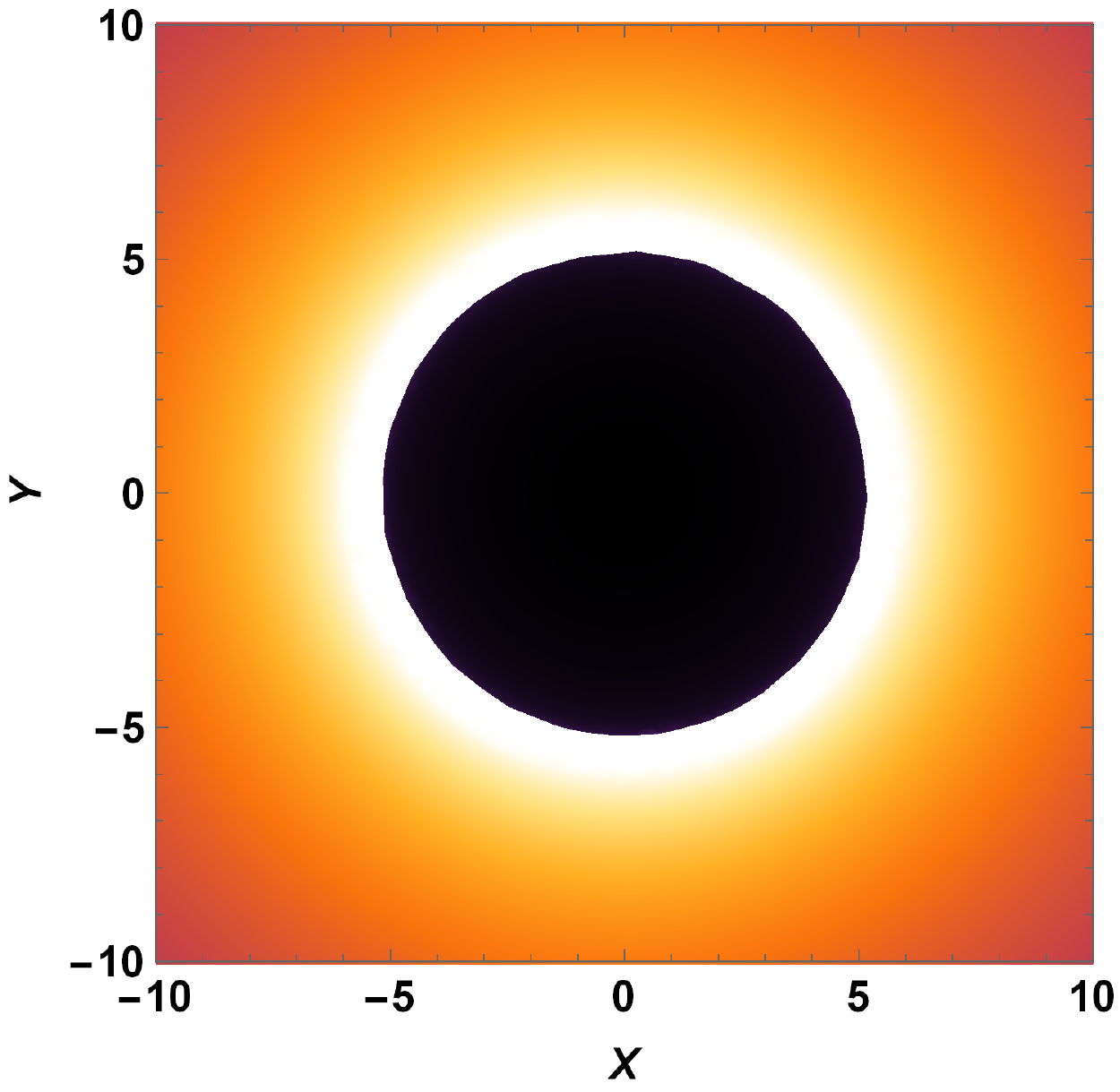}\label{re41}}\\
\subfigure[Intensity distribution in new spacetime.]
{\includegraphics[width=82mm]{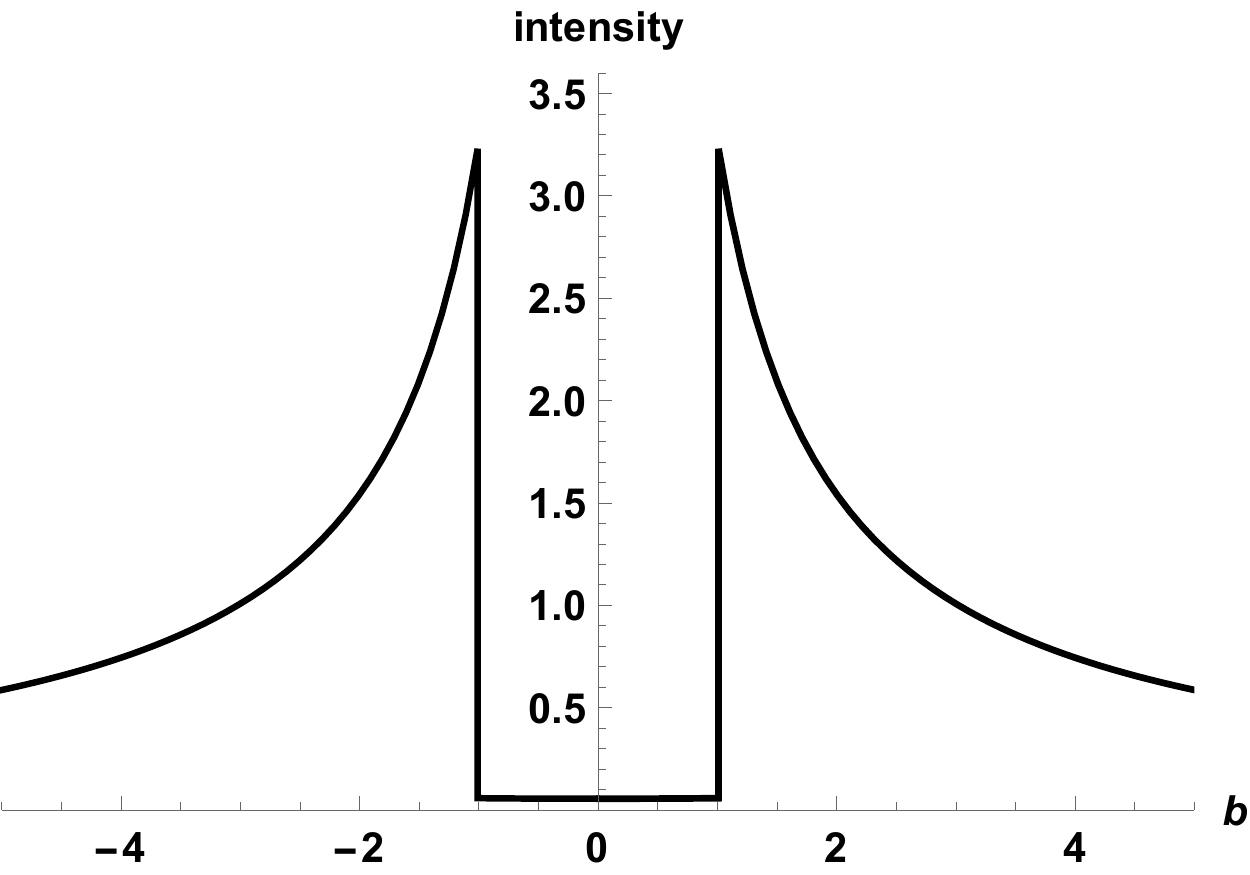}\label{re51}}
\hspace{0.2cm}
\subfigure[Shadow in new spacetime.]
{\includegraphics[width=68mm]{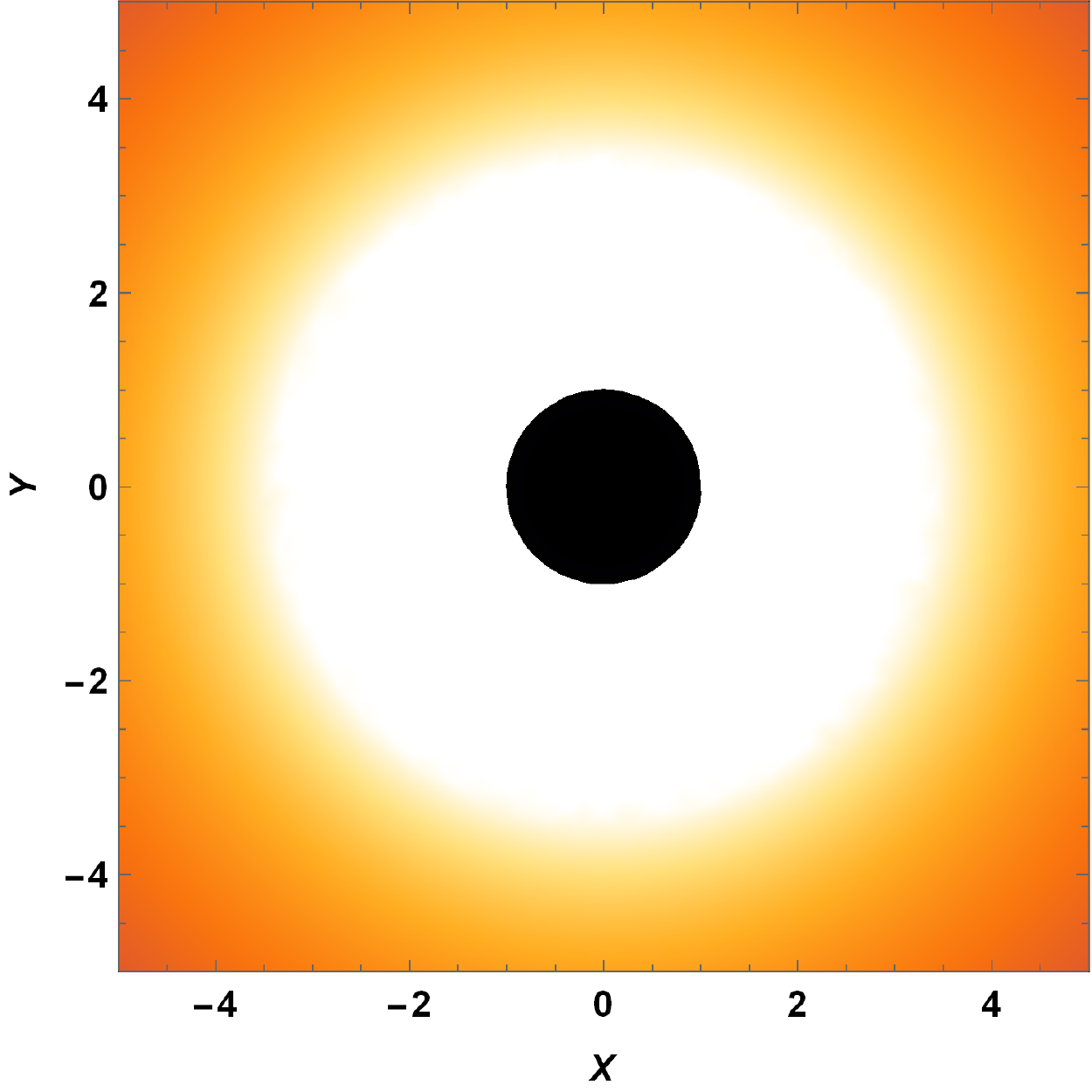}\label{re61}}
 \caption{In this figure, the intensity map in observer sky (left column figures) and the shadow of the central object (right column figures) are shown for Schwarzschild spacetime, JMN1 spacetime configuration (with $M_0=0.7$) and the naked singularity spacetime given here. The radius of the shadow of the Schwarzschild black hole (fig.~(\ref{re21})) and JMN1 naked singularity (fig.~(\ref{re41})) is $3\sqrt3$, where the Schwarzschild mass is $M_T=1$. The shadow shown in the right bottom corner is the shadow of the new naked singularity spacetime and the size of the shadow is $M=1$, where $M$ is the ADM mass.}
\label{fig2}
\end{figure*}

In this paper, for simplicity, we have considered spherically symmetric, radially freely falling, thin accreting matter which radiates monochromatic radiation, where the emissivity (in the emitter frame) per unit volume falls as,
\begin{equation}
    j(\nu_e) \propto \frac{\delta(\nu_e-\nu_*)}{r^2}\,\, ,
\end{equation} 
where $\nu_e$ is the emitted photon frequency  as measured in the rest frame of the emitter.
The intensity of the light coming out from the accreting matter can be written as the function of $X, Y$ which are the coordinates of the asymptotic observer's sky \cite{bambi}, 
\begin{equation}
    I_{\nu_o}(X,Y)= \int_{\gamma}^{} g^3 j(\nu_e) dl_{prop}\,\, ,\label{eq31}
\end{equation}
where $g=\nu_o/\nu_e$ is the redshift factor and $dl_{prop}$ is the infinitesimal proper length in the rest frame of emitter with $dl_{prop}=-k_\alpha u^\alpha_e d\lambda$, where $k^\mu$ is the null four velocity and $u^\alpha_e$ is the timelike four velocity of the emitter. Here $\lambda$ is the affine parameter and the integration is done along the lightlike trajectory $\gamma$. The redshift factor can be written as, 
\begin{equation}
    g=\frac{k_\alpha u^\alpha_o}{k_\beta u^\beta_e}
\end{equation}
where, $u^\mu_o = (1,0,0,0)$ is the four velocity of the static distant observer.
As we mentioned before, in this paper, we consider spherically symmetric accreting matter which is radially freely falling. The components of four velocity of radially freely falling particle in a general spherically symmetric, static spacetime (eq.~(\ref{static})) can be written as, 
\begin{equation}
    u^t_e = \frac{1}{g_{tt}},  u^r_e = -\sqrt{\frac{(1-g_{tt})}{g_{tt}g_{rr}}}, u^\theta_e = u^\phi_e = 0
\end{equation}
Using the components of four velocity of radially freely falling matter we can write down the redshift factor as,
\begin{equation}
    g = \frac{1}{\frac{1}{g_{tt}}-\frac{k_r}{k_t} \sqrt{\frac{(1-g_{tt})}{g_{tt}g_{rr}}}}\,\, ,
\end{equation}
where
\begin{equation}
    \frac{k^r}{k^t} = \sqrt{\frac{g_{tt}}{g_{rr}} \left(1-\frac{g_{tt} b^2}{r^2}\right)}\,\, .
\end{equation}
Now, the eq.~(\ref{eq31}) can be written as \cite{bambi},
\begin{equation}
    I_o(X,Y) \propto - \int_{\gamma}^{} \frac{g^3 k_t dr}{r^2 k^r}\,\, ,
\label{intensity}
\end{equation}
where $I_o(X,Y)$ is the intensity distribution in the (X, Y) plane of the observer's sky, where $X^2+Y^2=b^2$. Using the eq.~(\ref{intensity}), one can simulate the shadow. In figs.~ (\ref{re11}), (\ref{re31}), (\ref{re51}), we show how intensity varies with the impact parameter $b$ in the Schwarzschild, JMN1, and the naked singularity spacetime given here,  respectively. In figs.~ (\ref{re21}), (\ref{re41}), (\ref{re61}), we simulate the shadows cast by Schwarzschild, JMN1 and the new naked singularity spacetimes respectively.

From fig.~(\ref{fig2}), one can see that due to the presence of photon sphere in the Schwarzschild spacetime and in the external Schwarzschild spacetime of the JMN1 spacetime configuration (with $M_0>\frac23$), the shadow radius is $b_{ph}=3\sqrt3 M_T = 3\sqrt3$. The shadow cast by the JMN1 spacetime configuration with $M_0>\frac23$ is not distinguishable from the shadow cast by a Schwarzschild black hole  \cite{rajbul1}. On the other hand, the radius of the shadow  cast by the new naked singularity spacetime is $M$, where $M$ is the ADM mass of the spacetime. If we consider the ADM mass to be $M=1$, then the shadow radius will be $b_{tp}|_{r
_{tp}=0}=M=1$. Therefore, the shadow cast by Schwarzschild black hole and JMN1 naked singularity spacetime is $3\sqrt3$ times the size of the shadow cast by the naked singularity considered here.

As we know, in the near future, the Event Horizon Telescope (EHT) group will release the shadow of the Sagittarius A* (Sgr-A*). Therefore, it is very important to theoretically predict the size of the shadow. From the `S' stars motions around our galaxy center and from other physical phenomenon it is estimated that the mass of the central object of our milky way galaxy is around $4.31\times10^6 M_{\odot}$  with an error $\pm 0.38 M_{\odot}$ \cite{bambi}. If the central object is a black hole or JMN1 naked singularity then it is not possible to distinguish the black hole shadow from a naked singularity shadow. In table~(\ref{tabl1}), it is shown that the angular size of the shadow of  the Schwarzschild black hole and the JMN1 naked singularity is $56\pm 8~ \mu arcsec$. 

However, when such a spacetime exists where a naked singularity can form the shadow (e.g. the one we have analyzed here),  then the shadow size can be quite different for the same value of the ADM mass. In table~(\ref{tabl1}), we show that the naked singularity spacetime we discussed, with ADM mass $(4.31\pm 0.38)\times 10^6~ M_{\odot}$, would cast a shadow of diameter $10\pm 8~\mu arcsec$. 

   \begin{table}
    \centering
    \caption{\textbf{Mass of Sagittarius A* and its Shadow size in Naked singularity and Black hole spacetimes}}
    \begin{tabular}{l c c}
        \hline\hline
Object   &   Mass   &   Shadow (diameter) \\
     &   ($M_{\odot}$)   &   ($\mu arc sec$) \\
\hline
Black hole  &    $(4.31\pm 0.38)\times10^6$   &  $56\pm8$ \\
JMN Naked singularity  &     $(4.31\pm 0.38)\times10^6$  & $56\pm 8$ \\
New Naked singularity  &    $(4.31\pm 0.38)\times10^6$  & $10\pm 8$ \\
\hline
    \end{tabular}
    \label{tabl1}
\end{table}
\vspace{1cm}

\section{Conclusion}\label{result}
In this paper, we study the nature of light geodesics and shadow in a new  naked singularity spacetime, and compare the results with the light geodesics and shadow in the Schwarzschild and JMN1 (with $~M_0>\frac23$) spacetimes. We show that the nature of light geodesics in JMN1 and Schwarzschild spacetimes are quite different from the light geodesics in the proposed naked singularity spacetime. We also show that the shadow cast by the photon sphere (Schwarzschild and JMN1 with $M_0>\frac23$ case) is $3\sqrt3$ times bigger than the shadow cast by the singularity, that happens for the new naked singularity spacetime, when we take same ADM mass for all spacetimes. 

The main purpose of this work is to show that the shadow is not only the property of a black hole or a photon sphere, but a singularity by itself can also cast a shadow, as demonstrated explicitly by the example spacetime we have considered and proposed here. 
An important point that follows is, one of the observable signatures of a naked singularity is the shadow that it could cast. Thus such a shadow could be an important property of a naked singularity.  Further, the important characteristic of such a shadow of singularity is, it would be considerably smaller in size, as compared to the shadow of a black hole. This would provide an important characteristic and observable difference that might distinguish a singularity from the black hole.  

Near the naked singularity, the quantum gravity effects should be dominant, and therefore, such quantum gravity effects might be menifested or can be observed in the shadow cast by a naked singularity. This will require a detailed analysis of the various features encoded in such shadows. Hence, this novel feature of a naked singularity may be  important as well in the context of recent observations of the shadow of the galactic center M87 \cite{Akiyama:2019fyp}, and for the upcoming image of the center of Milky-way galactic center (Sgr-A*).  
Of course, what we have given here is a case study, and more detailed analysis will be needed to know if shadows are a general feature associated with singularities.

It is intriguing to note that when proposed initially, black holes were considered to be highly exotic objects, and their existence was even doubted sometimes. The past few decades have, however, seen them to emerge as more realistic astrophysical objects, used to explain several cosmic phenomena. Similarly, spacetime singularities, and the naked singularities are seen as probable exotic entities in the cosmos. But, as shown here, if they manifest observable features such as casting and creating shadows in their own right, this could be an interesting and useful step to bring them in an observable domain and in the realm of physical reality.

%Event horizon telescope(EHT) realise image of M87 and also will realise Sgr. A* in near future. But shadow did not give the correct information about the spacetime in the galactic centre. For that, experimentally black hole in our Milkey way galaxy only conform when two or more then two S stars precession perfectly match and by using this get exact same mass value, and this mass value also conform by light trajectories, like using lensing and shadow of the central object.\\
%In given new spacetime is very interesting, its approximation is Schwartzschild spacetime and event horizon is lies in $r=0$ and physical strong singularity also lies in $r=0$. So we can say that, this can be null strong singularity at the centre. In near future we can study gravitational collapse of this spacetime and accretion disk properties of spacetime.\\
%Main motivation of the paper is to show the example in which naked singularity cast shadow without any photon sphere.
 
%%%%%%%%%%%%%%%%%%%%%%%%%%%%%%%%%%%%%%%%%%%%%%%
%%%%%%%%%%%%%%%%%%%%%%%%%%%%%%%%%%%%%%%%%%%%%%%%

\end{document}